\documentclass[fleqn,usenatbib]{mnras}

\usepackage{newtxtext,newtxmath}

\usepackage[T1]{fontenc}
\usepackage{ae,aecompl}

\usepackage{graphicx}	
\usepackage{amsmath}	

\usepackage{gensymb}

\title[Searching MPs in Rup106]{Searching for Multiple Populations in Ruprecht 106}

\author[Frelijj et al.]{
H. Frelijj,$^{1}$\thanks{E-mail: hfrelijj@astro-udec.cl}
S. Villanova,$^{1}$
C. Mu\~noz,$^{1,2,3}$
J. G. Fern\'andez-Trincado$^{4,5,6}$
\\
$^{1}$Departamento de Astronom\'ia,  Casilla 160-C, Universidad de Concepci\'on, Chile\\
$^{2}$Departamento de Astronom\'ia, Universidad de La Serena, Av. Juan Cisternas 1200, La Serena,Chile\\
$^{3}$Instituto de Investigaci\'on Multidisciplinario en Ciencia y Tecnolog\'ia, Universidad de La Serena. Avenida Ra\'ul Bitr\'an S/N,\\
La Serena, Chile.\\
$^{4}$Instituto de Astronom\'ia y Ciencias Planetarias, Universidad de Atacama, Copayapu 485, Copiap\'o, Chile\\
$^{5}$Institut Utinam, CNRS UMR 6213, Universit\'e Bourgogne-Franche-Comt\'e, OSU THETA Franche-Comt\'e, Observatoire de Besan\c{c}on,\\
BP 1615, 25010 Besan\c{c}on Cedex, France\\
$^{6}$Centro de Investigaci\'on en Astronom\'ia, Universidad Bernardo O'Higgins, Avenida Viel 1497, Santiago, Chile
}

\date{Accepted 2021 February 10. Received 2021 February 8; in original form 2021 January 1}

\pubyear{2021}

\begin{document}
\label{firstpage}
\pagerange{\pageref{firstpage}--\pageref{lastpage}}
\maketitle

\begin{abstract}
More than a decade has passed since the definition of Globular Cluster (GC) changed, and now we know that they host Multiple Populations (MPs). But few GCs do not share that behaviour and Ruprecht 106 is one of these clusters. We analyzed thirteen member red giant branch stars using spectra in the wavelength range 6120-6405 \AA \ obtained through the GIRAFFE Spectrograph, mounted at UT2 telescope at Paranal, as well as the whole cluster using C, V, R and I photometry obtained through the Swope telescope at Las Campanas. Atmospheric parameters were determined from the photometry to determine Fe and Na abundances. A photometric analysis searching for MPs was also carried out. Both analyses confirm that Ruprecht 106 is indeed one on the few GCs to host Simple Stellar Population, in agreement with previous studies. Finally, a dynamical study concerning its orbits was carried out to analyze the possible extra galactic origin of the Cluster. The orbital integration indicates that this GC belongs to the inner halo, while an Energy plane shows that it cannot be accurately associated with any known extragalactic progenitor.
\end{abstract}



\section{Introduction}
  The  classical  paradigm  of  Galactic  globular  clusters  (GCs)  being  simple stellar  populations  has  changed  dramatically  by  observational  evidence.  The  presence  of  chemical  inhomogeneities  in its light elements, like a spread in Na or O extently studied by \citet{Carretta2009}, led some authors to create different theories to try to explain this behaviour \citep{Dercole2008,Renzini2015,Bastian2013}, but to the date no one satisfies all the observations.
 As mentioned, the abundance analysis (obtained through spectroscopy) like the Na-O, Mg-Al and C-N anticorrelation is one of the strongest weapons to determine whether a GC posses MPs since different populations present different chemical abundances.
 
 Photometry is a different approach to search for MPs when chemical abundances are not available. We can use it to analyze large groups of stars simultaneously since UV filters have proved to be sensitives to differences in chemical abundances of light elements. \citet{Bedin2004} used HST observations to prove that using the right combination of filters it is possible to distinguish between multiple main sequences and/or sub-giant and/or giant branches in almost all GCs. The most popular works involving this technique was \citet{Piotto2015} who used an improved combination of the UV/blue WFC3/UVIS filters F275W, F336W and F438W, the so called 'magic trio', to analyze ~60 GCs. All of them presented splitted CMD sequences or, at least, a great broadening in some parts of the sequences. Other filters like the C filter from the Washington Filter system \citep{Canterna1976} have proved to be very sensitive in discriminating the presence of MPs \citep{Cummings2014,Frelijj2017}. But like most cases, there seem to be an exception to the rule. A bunch of GCs seems to host a Simple Stellar Population(SSP)(i.e E3 \citep{Salinas2015}, Terzan 7 \citep{Tautvaivien2004}, Ruprecht 106 \citep{Villanova2013}, etc), although actually some of them do not have enough evidence to confirm this hypothesis. In this paper we are going to focus in one of these: Ruprecht 106 (hereafter Rup106).

According to \citet[2010 edition]{Harris1996} Rup106 is a GC with $\alpha$(J2000): 12$^h$38$^m$40.2$^s$ and $\delta$(J2000): -51$\degree$09'01'', located at 21.2 kpc from the Sun and 18.5 kpc from the Galactic centre. It has a metallicity [Fe/H] = -1.68, an Heliocentric Radial Velocity RV = -44 $\pm$ 3 km s$^{-1}$ and a Foreground Reddening E(B-V) = 0.2.

As mentioned, \citet[hereafter V13]{Villanova2013} realized a spectroscopic study in Rup106 showing that all the 9 samples studied didn't show a Na-O anticorrelation, concluding that Rup106 is one of the few GCs that lacks to exhibit the phenomenon of MPs, which have been well supported by an independent   photometric study by \citet{Dotter2018}. V13 also show that Rup106 has an extragalactic origin since its very low Na and $\alpha$-element abundances only match those of the Magellanic Clouds and of the Sagittarius Galaxy. Both studies together present strong evidence of Rup106 host a SSP, but the 9 targets from V13 are not enough to assure that the cluster has no chemical spread at all. This paper tries to provides, along with V13, enough spectroscopic and photometric evidence to demonstrate that Rup106 is indeed a SSP cluster or that possess at least one star from a different population, what could open again the controversy about this cluster.

 This paper is organized in this way: In section 2, we discuss our observations and data reduction from Photometry and Spectroscopy. In section 3 we detail the steps done to get Heliocentric Velocities and Proper Motions to filter our photometric catalogue from non members. We also describe the process we used to get the atmospheric parameters that are necessary to calculate abundances. Section 4 describes the abundance determination. Section 5 presents the photometric analysis we applied to determine either Rup106 have MPs or not. Section 6 contains a study of the orbits discussing the possible extragalactic origin of Rup106. Section 7 present a resume of our results.


\section{Data}
\subsection{Photometric Observations}
The sample used for this work is composed of 21 photometric images taken at the 1-m Swope telescope, Las Campanas Observatory, Chile. All of these images are from the same observing run in march 2014. 
 The Swope telescope works with one CCD (E2V CCD231-84) which contains 4 amplifiers together forming a square of 4096x4112 pixels with a scale of 0.435 ''/pix and a field of view of 29.7x29.8 arc minutes.
 The filters used for this work are the Washington C filter \citep{Canterna1976}, the I$_{\text{KC}}$ filter, and the Harris V and R filters.
  From the 21 images, 2 images were taken using the I filter, 5 using R, 6 images using V and 8 C.
  
  Table \ref{tab:exposures} Details of the exposures:
  
{%
\newcommand{\mc}[3]{\multicolumn{#1}{#2}{#3}}
\begin{center}
\label{tab:exposures}
\begin{tabular}{|l|l|}\hline
 & \mc{1}{c|}{N\degree (exposure time)}\\\hline
C & 1(30s), 1(300s), 6(1200s)\\
V & 2(10s), 1(100s), 3(400s)\\ 
R & 1(10s), 1(100s), 3(400s)\\
I & 1(10s), 1(100s)\\\hline
\end{tabular}
\end{center}
}%

The FWHM of the images ranges between 1.37''-2.15'' and the airmasses vary between 1.079-1.160. None of the nights were considered hotometric, hence standard fields could not be observed.

The data processing and reduction was performed according to \citet{Frelijj2017}.
We used \textsc{iraf} \footnote{\textsc{iraf} is distributed by the National Optical Astronomy Observatory, which is operated by the Association of Universities for Research in Astronomy (AURA) under a cooperative agreement with the National Science Foundation.}
to process all the 4 quadrants of each image separately, more specifically its tasks ccdproc, zerocombine and flatcombine. After that a script was used to combine the four quadrants into one single image.
The photometry was performed using \textsc{daophot} \citep{Stetson1987} since this program is the indicated to threat crowded fields. We obtained the PSF using the brightest non saturated and isolated stars in each frame. With a good PSF at hand we proceeded to run \textsc{allstar} in each image separately. Once finished, we aligned all the catalogues with \textsc{daomatch} and \textsc{daomaster}. The file with the transformation coordinates was used along with the images, the catalogs and psf files to finally run \textsc{allframe} \citep{Stetson1994}. \textsc{allframe} made PSF-photometry simultaneously in all the frames to realize the best photometry.
Finally, with each image catalogue that \textsc{allframe} returned, we apply aperture corrections realizing aperture photometry to the PSF-Stars and comparing it with the PSF-photometry.

\begin{figure}
\includegraphics[width=0.5\textwidth]{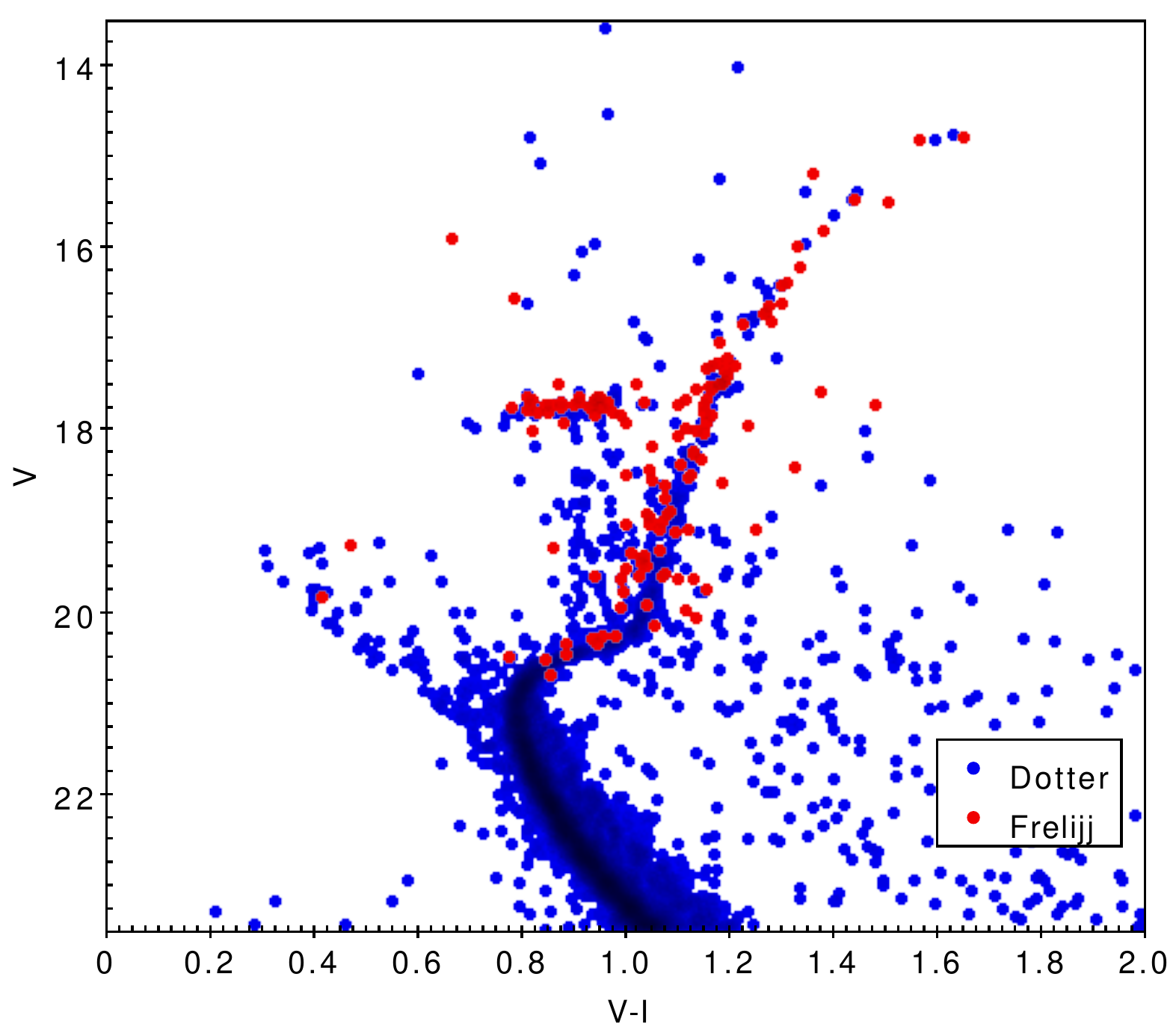}
\caption{\label{Figure comp}Plot overlapping a CMD using V$_{\text{ground}}$-I$_{\text{ground}}$ vs V$_{\text{ground}}$ from \citet{Dotter2011} and CMD with our calibrated V-I vs V.}
\end{figure}

We calibrated our data using the Catalog used in \citet{Dotter2011} available in the ACS GC Treasury database, where they took the filters F606W and F814W and converted them to ground-based filters V and I (hereafter V$_{\text{ground}}$ and I$_{\text{ground}}$) using the relation from \citet{Sirianni2005}. After matching them with our catalogue we derived the transformation equations in the form of:$$V = (v-i)*m + b + v$$
$$I = (v-i)*m + b + i$$
 Where V and I are our calibrated magnitudes, m is the slope, b is the y-intercept of the line and v and i our instrumental magnitudes. To verify the accuracy of the calibration we calculated the difference V$_{\text{ground}}$-V and I$_{\text{ground}}$-I. For the I filter we found a residual shift of 0.04 mag., so we subtracted it to all the calibrated I magnitudes.
 Figure \ref{Figure comp} shows a comparison between the V$_{\text{ground}}$-I$_{\text{ground}}$ vs V$_{\text{ground}}$ CMD from the \citet{Dotter2011} catalogue (blue dots) and our calibrated catalogue (red dots), both HB and RGB are aligned, probing that the calibration is fine. We could not find a way to calibrate C and R filters, but for this it was not necessary.

\citet{Bonatto2013} shows that this cluster has a mean differential reddening of $\langle$E(B-V)$\rangle$ = 0.026 $\pm$ 0.010 with a maximum differential reddening of $\delta$E(B-V)$_{\text{max}}$ = 0.051, indicating that we do not need to make differential reddening corrections.

Finally, the x/y coordinates of the standardized catalog were transformed to RA/Dec(J2000) using the xy2sky task from \textsc{wcstools} and a World Coordinate System created using 10 well separated stars from the reference frame with the \textsc{iraf} tasks ccmap and ccsetwcs.


\subsection{Spectroscopic Observations}
\begin{figure}
\includegraphics[width=0.5\textwidth]{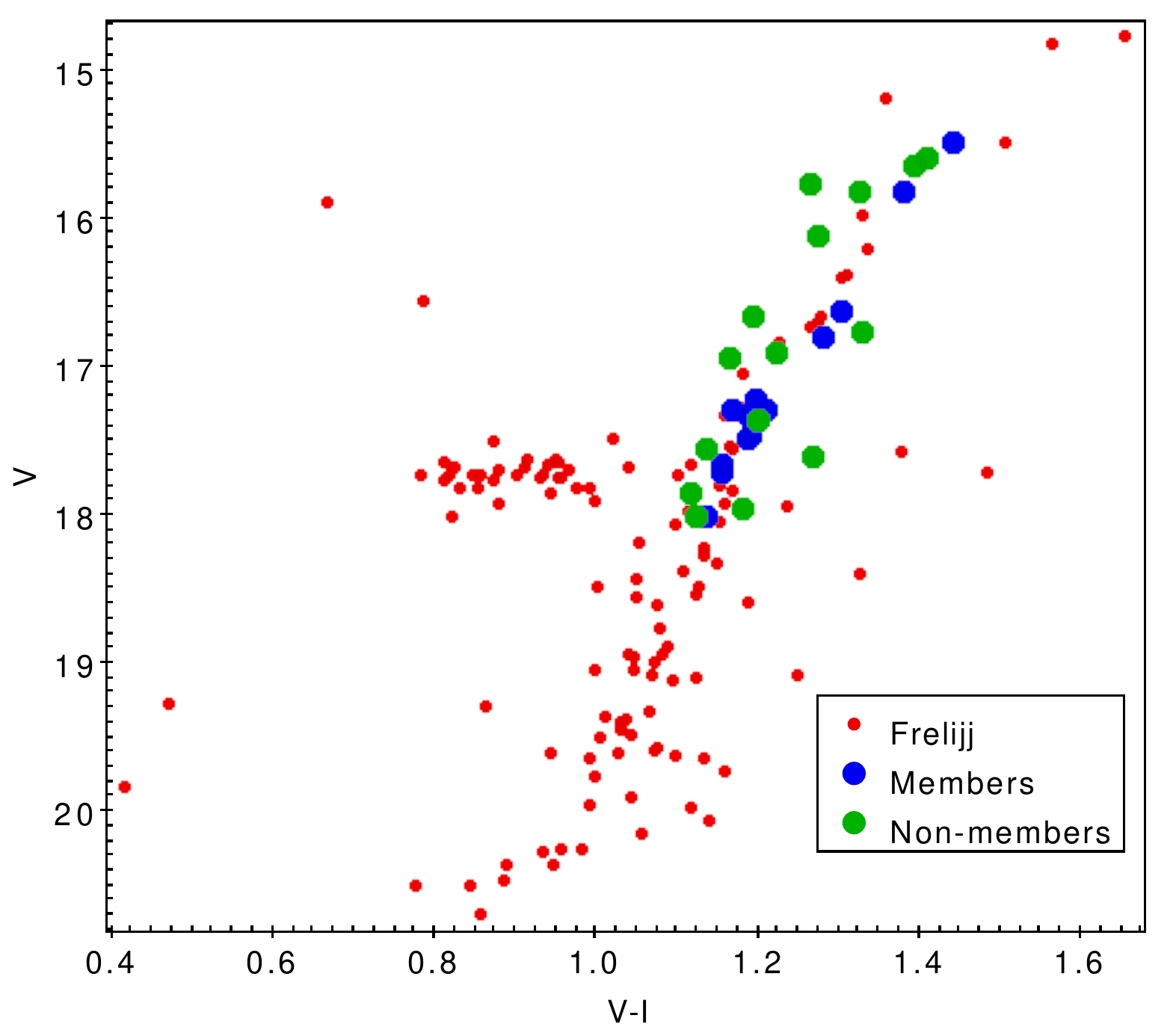}
\includegraphics[width=0.5\textwidth]{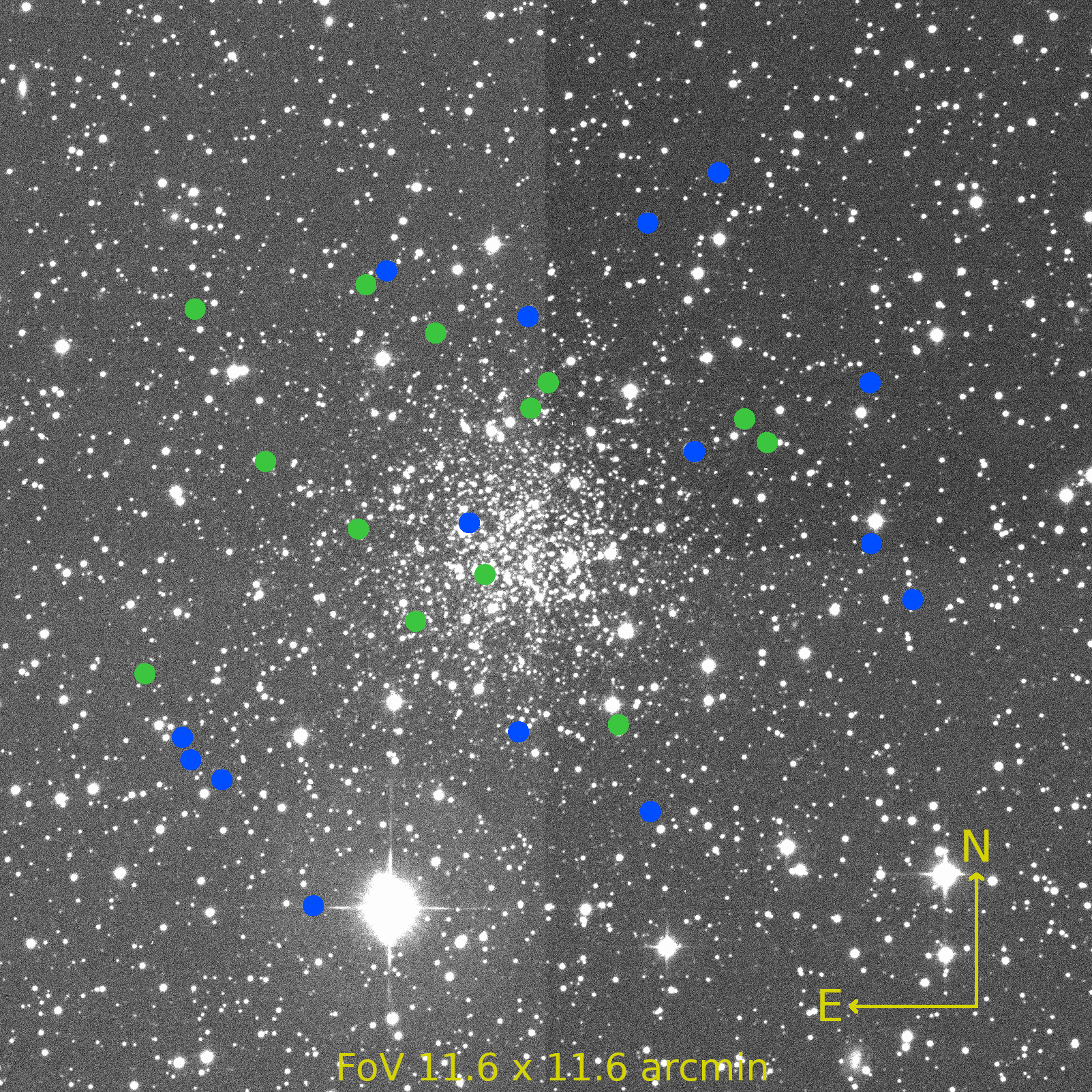}
\caption{\label{fulltarg}
Top: Cluster Targets have been highlighted in the CMD. Members are shown as Green dots while non-members are shown as Blue dots. Bottom: Spatial distribution of the same targets along the cluster.}
\end{figure}

 Our spectroscopic data consist in observations from 2017 as part of the programme ID 098.D-0227(A) obtained using the medium-high resolution FLAMES-GIRAFFE Spectrograph installed in the UT2 (Kueyen) telescope in Paranal. The resolving power is R$\sim$26400. Our targets were selected in the magnitude range V = 15.5 and V = 18.5 and they belong to the RGB (Figure  \ref{fulltarg}). 
 We observed 28 stars in the Wavelength range 6120-6405 \AA. The exposure time was 2640 seconds per spectrum, and each star was observed 4 times, getting a total of 112 spectra.
 
 The spectroscopic data were reduced using the GIRAFFE pipeline, with only a normalization, sky subtraction and a transformation from nm to \AA \ remaining to do. These steps were done using \textsc{iraf} tasks, specifically continuum, sarith and hedit. The 4 spectra of each star were combined using the task scombine to improve the S/N ratio.

 We measured observed radial Velocities using the \textsc{iraf} fxcor package, with the help of a synthetic spectrum as a template calculated using typical RGB star parameters, i.e. T$_{\text{eff}}$ = 4500 K, log(g) = 1.50, v$_\text{t}$ = 1.50 km s$^{\text{-1}}$, and the metallicity of the cluster ([Fe/H] = -1.50). These relative velocities were used to apply Doppler corrections through the \textsc{iraf} task dopcor.


\section{Data Analysis}
\subsection{Heliocentric Radial Velocities, Proper motions and membership}

\begin{figure}
\includegraphics[width=0.5\textwidth]{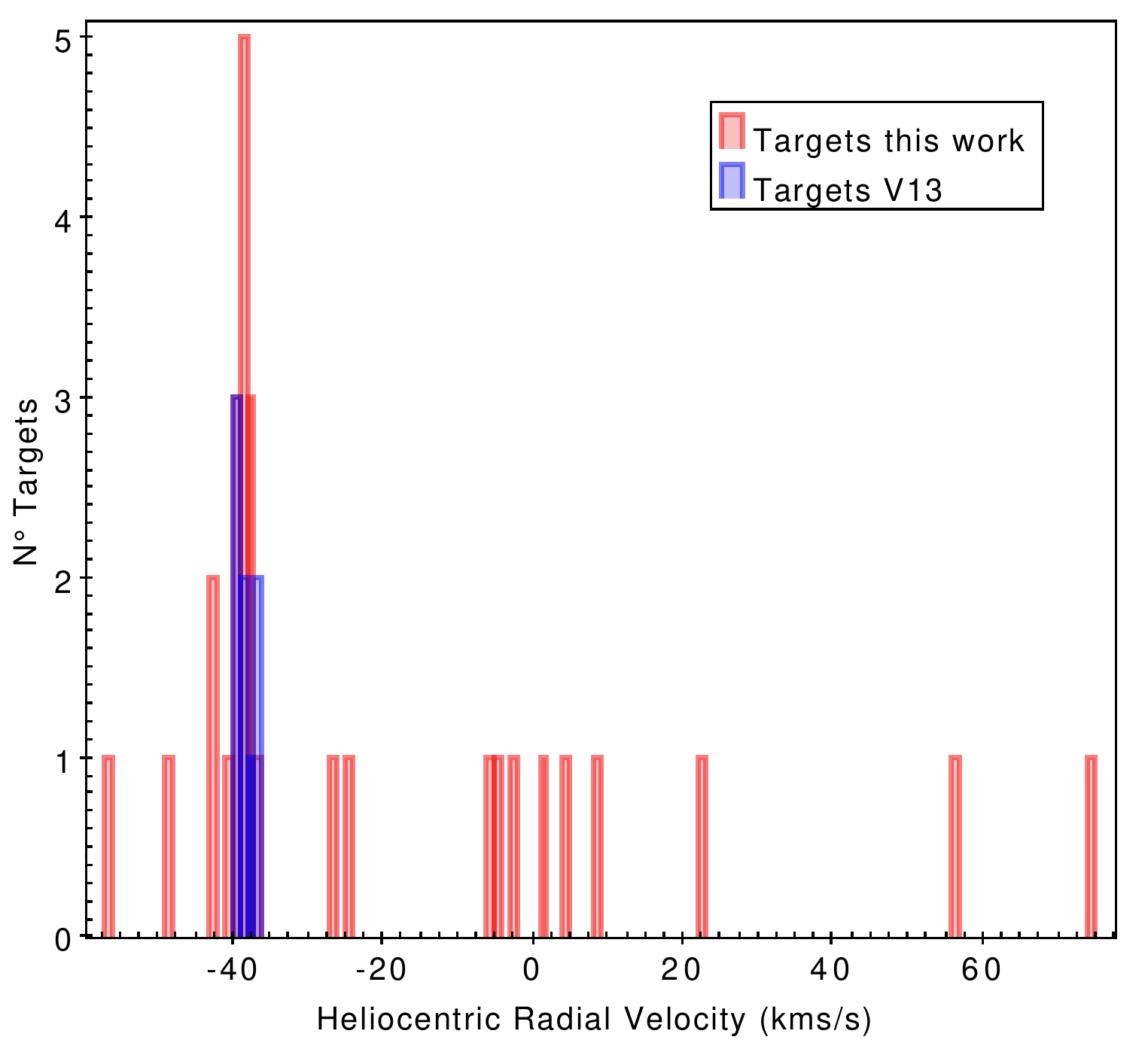}
\caption{\label{rvcomp}Heliocentric Velocities of the targets of this study are represented in red, targets from \citet{Villanova2013} have been overlapped in blue. All targets between -45 km s$^{\text{-1}}$ and -35 km s$^{\text{-1}}$ are considered members of the cluster.}
\end{figure}

We obtained heliocentric radial velocities (RV) using the \textsc{iraf} task rvcorrect. According to Figure  \ref{rvcomp}, we consider possible cluster members those targets with RV between -45 and -35 km s$^{-1}$. This reduced our targets to 15. The mean heliocentric velocity of this sample is -38.99 $\pm$ 1.6 km s$^{-1}$ with a standard deviation of 1.7 km s$^{\text{-1}}$. These values are in good agreement with V13 as it is shown in Figure  \ref{rvcomp}.

\begin{figure}
\includegraphics[width=0.5\textwidth]{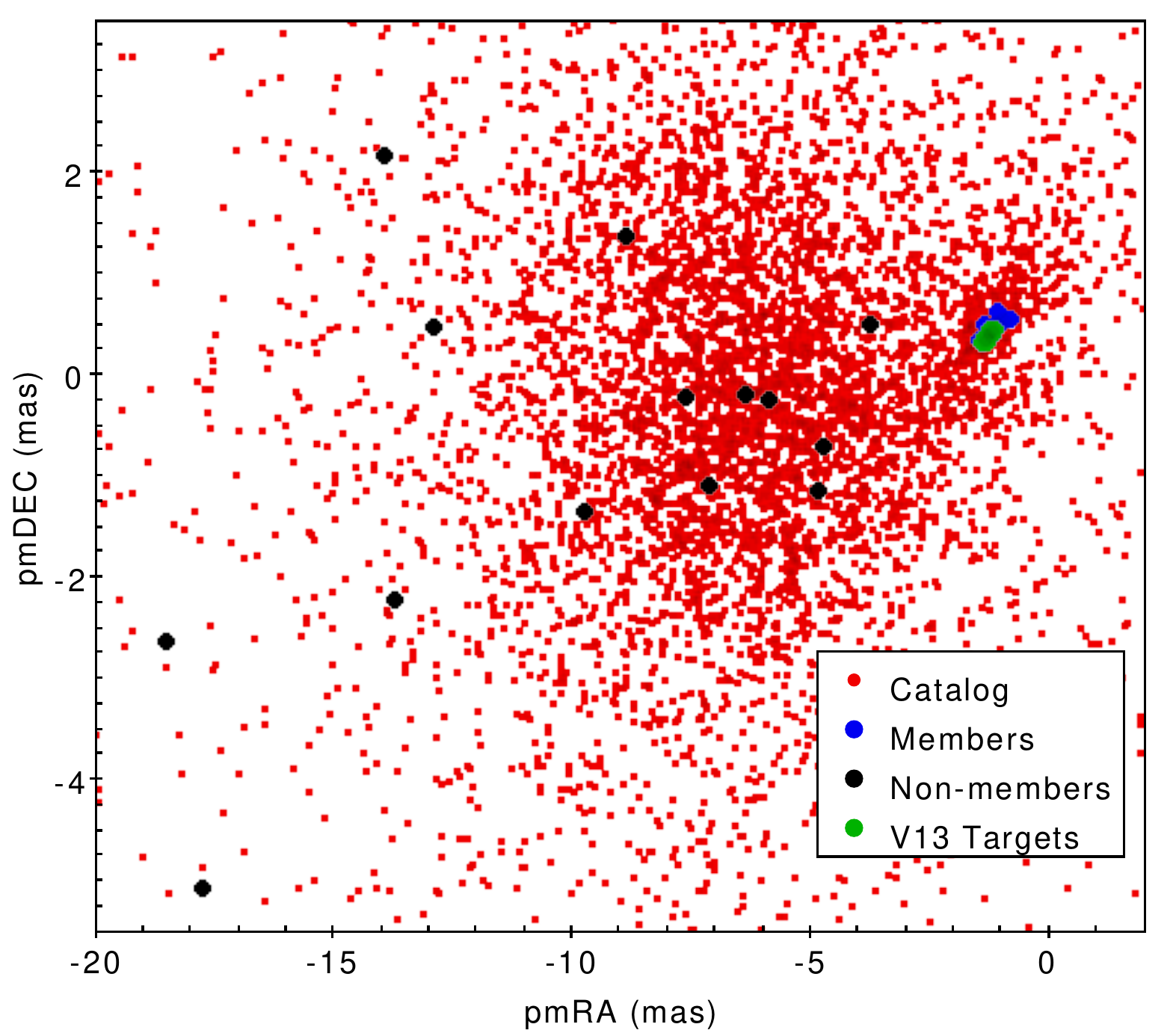}
\caption{\label{pm}Proper motions for each star of the catalog. Blue dots represent the 13 targets members of the Cluster, black dots represent discarded non-members targets, and green dots represent the targets from V13. }
\end{figure}

Thanks to the Proper Motions (PM) provided by the \textit{Gaia} mission \citep{Gaia2016,Gaia2018}, we could remove further non-member stars as shown in Figure  \ref{pm}. We discarded 2 more of our targets that had RV similar to the mean radial velocity of the cluster but very different PM. The average PM of our targets are : pmRA = -1.21 $\pm$ 0.13 mas yr$^{\text{-1}}$ and pmDEC = 0.43 $\pm$ 0.08 mas yr$^{\text{-1}}$ \footnote{In agreement with \textit{Gaia} EDR3\citep{Gaia2020}.}.
Table \ref{coord} lists the details of the final members. 


\subsection{Atmospheric parameters}
\begin{figure}
\includegraphics[width=0.517\textwidth]{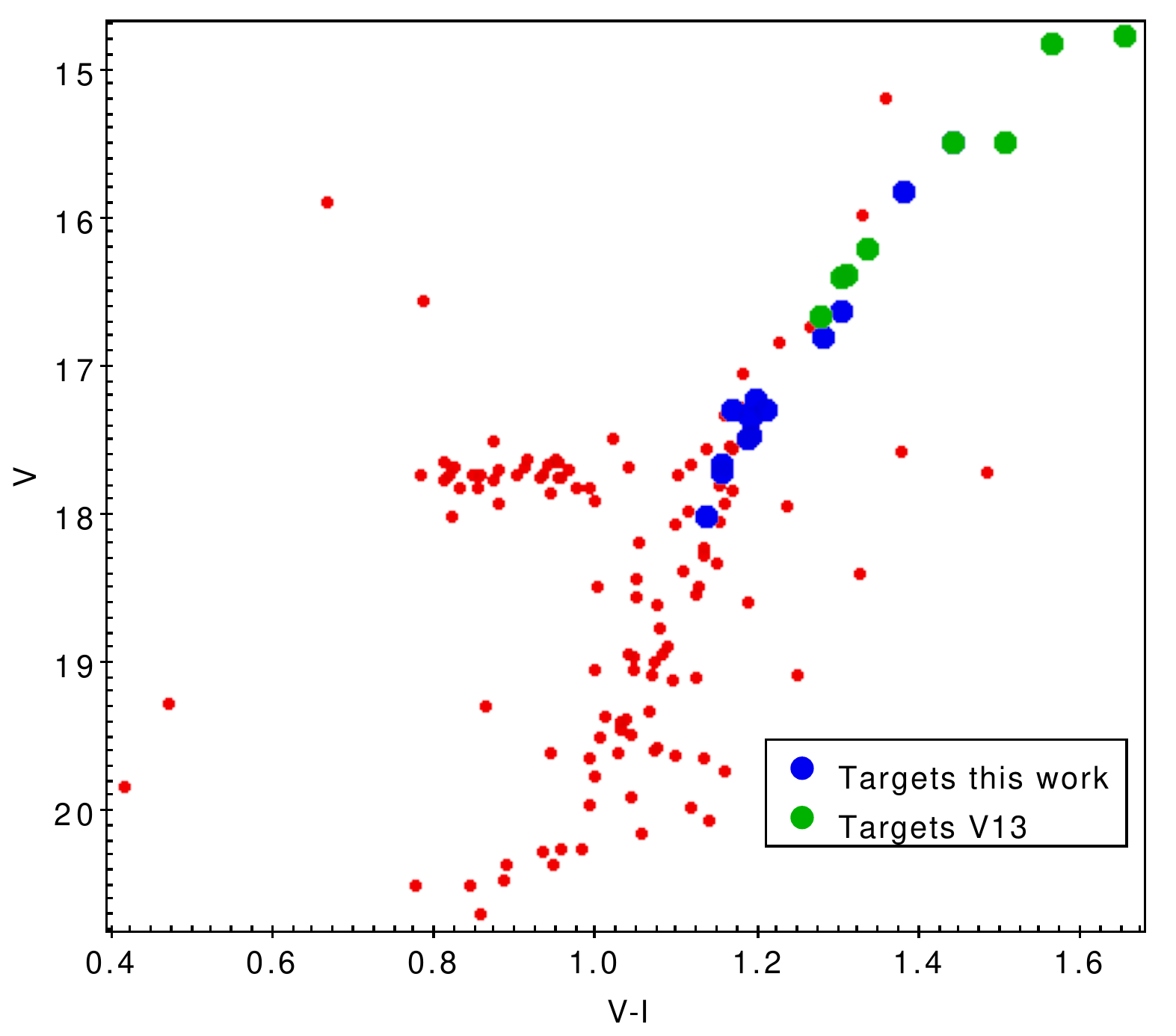}
\caption{\label{V-I}Definitive CMD, targets from this work appear as blue dots while targets from V13 are the green dots.}
\end{figure}

Figure \ref{V-I} shows the V-I vs V CMD with the identified members from this work and V13. All the stars with photometric errors greater than 0.1 were removed.

\begin{figure}
\includegraphics[width=0.5\textwidth]{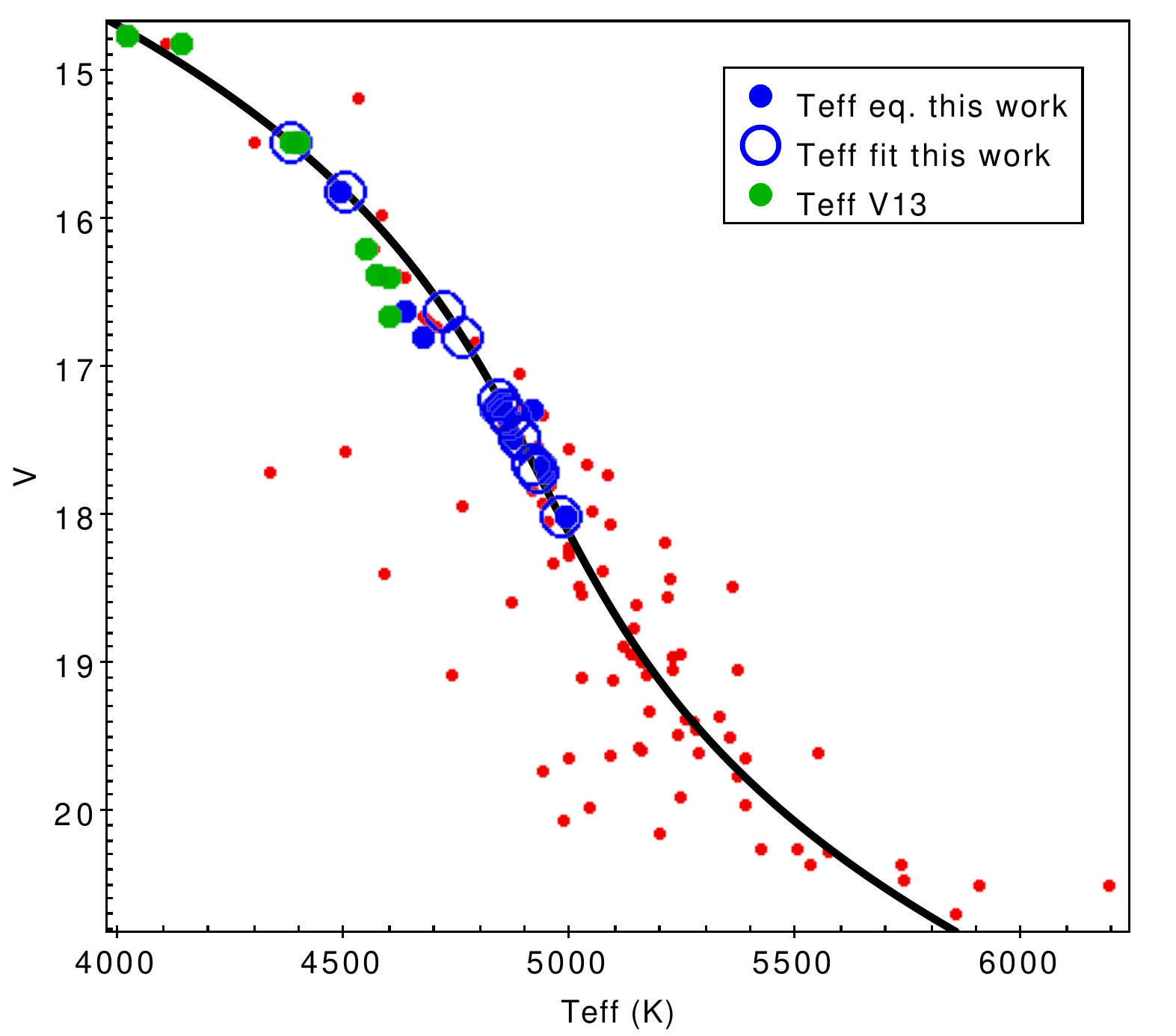}
\caption{\label{poly}effective temperatures for each RGB star are shown. Average value from \citet{Alonso1999} and \citet{Ramirez&Melendez2005} appear as red dots. A polynomial fit is shown in black, filled blue dots indicate targets from this work obtained through the mentioned relations while blue circles are the values obtained through the polynomial fit, green dots are the effective temperatures from V13.}
\end{figure}

A reddening correction E(B-V) = 0.20 \citep[2010 edition]{Harris1996} was applied to the V-I color using the extinction relation E(B-V) = 1.24E(V-I) in order to obtain effective temperatures. Then T$_{\text{eff}}$ were determined averaging the values obtained from the expressions given in \citet{Ramirez&Melendez2005} and \citet{Alonso1999}. Since \citet{Alonso1999} works with Johnson colors, a relation  $$(V-I)_J=-0.005 + 1.273\cdot(V-I)_C$$ from \citet{Fernie1983} was applied to our V-I color.
Then T$_{\text{eff}}$ were plotted against V and a polynomial was adjusted to the RGB, making possible to obtain T$_{\text{eff}}$ using the V magnitudes instead of V-I, reducing the uncertainties. These values obtained through the polynomial fit are our definitive T$_{\text{eff}}$. Figure  \ref{poly} shows a comparison between the T$_{\text{eff}}$ obtained through the formulae and the polynomial,  T$_{\text{eff}}$ from V13 are shown for comparison.

Surface gravities log(g) were determined through the canonical equation: $$log(g/g_{\odot}) = 4log(T_{eff}/T_{\odot}) - log(L/L_{\odot}) + log(M/M_{\odot})$$
Assuming a mass of 0.8 M$_{\odot}$, a luminosity based in the distance modulus (m-M)$_\text{v}$ = 17.25 \citep[2010 edition]{Harris1996} and a relation obtained from \citet{Alonso1999} for bolometric corrections(BC).
Finally, microturbulence velocities v$_\text{t}$ were determined using the relation from \citet{Gratton1996}:
$$v_t=2.22-0.322\cdot log(g)$$

Table \ref{coord} gives the values of the atmospheric parameters.

\begin{table*}
\centering
 \begin{tabular}{c c c c c c c c c c c} 
 \hline
  ID & RA(J2000) & Dec & V & I & pmRA & pmDEC & RV & T$_{\text{eff}}$ & log(g) & v$_\text{t}$\\
  & (h:m:s) & ($\degree$:':'') & mag & mag & (mas yr$^{\text{-1}}$) & (mas yr$^{\text{-1}}$) & (km s$^{\text{-1}}$) & (K) & (dex) & (km s$^{\text{-1}}$)\\
  \hline
  11012 & 12:38:33.50 & -51:10:58.30 & 18.022 & 16.885 & -1.47079 & 0.34399 & -39.16 & 4978 & 2.3715  &   1.4564\\
  11579 & 12:39:05.60 & -51:10:26.60 & 17.303 & 16.093 & -1.24388 & 0.38164 & -37.88 & 4856 & 2.0253  &   1.5678\\ 
  12911 & 12:38:42.58 & -51:09:23.00 & 15.494 & 14.054 & -1.37353 & 0.35648 & -39.25 & 4382 & 1.0307  &   1.8881\\
  14650 & 12:38:57.47 & -51:08:11.40 & 17.231 & 16.033 & -1.07649 & 0.60763 & -40.31 & 4843 & 1.9899  &   1.5792\\
  14861 & 12:38:23.50 & -51:07:58.40 & 17.484 & 16.292 & -1.4267  & 0.31629 & -38.29 & 4888 & 2.1132  &   1.5396\\
  15108 & 12:38:25.04 & -51:07:43.40 & 17.502 & 16.315 & -1.26708 & 0.40051 & -42.77 & 4891 & 2.1220  &   1.5367 \\
  15225 & 12:38:39.52 & -51:07:37.10 & 17.673 & 16.517 & -1.3498  & 0.48873 & -39.50 & 4920 & 2.2042  &   1.5103  \\
  15502 & 12:38:38.34 & -51:07:20.70 & 16.807 & 15.526 & -1.18759 & 0.38536 & -37.29 & 4761 & 1.7783  &   1.6474 \\
  15985 & 12:38:45.98 & -51:06:49.40 & 16.631 & 15.328 & -1.12986 & 0.47452 & -38.71 & 4723 & 1.6877  &   1.6766\\
  16174 & 12:39:02.27 & -51:06:34.50 & 17.725 & 16.570  & -0.94386 & 0.50756 & -38.09 & 4928 & 2.2292  &   1.5022\\
  16394 & 12:38:50.70 & -51:06:18.80 & 15.822 & 14.439 & -1.09852 & 0.46612 & -38.02 & 4501 & 1.2368  &   1.8218\\
  5015399 & 12:38:47.27 & -51:09:52.90 & 17.351 & 16.159 & -0.82873 & 0.54844 & -37.87 & 4865& 2.0487 &    1.5603\\
  5016747 & 12:38:51.17 & -51:08:54.20 & 17.305 & 16.137 & -1.27574 & 0.34161 & -42.74 & 4856& 2.0264 &    1.5675\\

 \hline
\end{tabular}
\caption{Table with the details of the Targets. The order of the columns are: Star ID, Right Ascension(J2000), Declination(J2000) , Magnitude in V, Magnitude in I, Absolute Proper Motion in RA, Absolute Proper Motion in DEC, Heliocentric Radial Velocity, effective temperature, surface gravity and microturbulence velocity.}
\label{coord}
\end{table*}

The T$_{\text{eff}}$, log(g) \& v$_\text{t}$ were used together with the metallicity [Fe/H] = -1.5 (V13) to generate atmospheric models for each target.



\section{Abundance Analysis}
Chemical abundances were calculated using the Local Thermodynamic Equilibrium program \textsc{moog} \citep{Sneden1973}, and atmospheric models were calculated using the \textsc{atlas9} code \citep{Kurucz1970}, assuming our initial estimations of the atmospheric parameters.
 The Spectrum-synthesis technique was used to determine Fe and Na abundances. This method consist in comparing an observed spectral line with five different synthetic spectra calculated with different abundances. The interpolated model with the lowest Root Mean Square give us the abundance of the element associated with that line. For a more precise determination we applied a parabolic fit to the 5 RMS values of the 5 synthetic spectra plotted as a function of the abundance in order to obtain the minimum. This minimum is the final abundance we assumed for the line.

\begin{figure}
\includegraphics[width=0.515\textwidth]{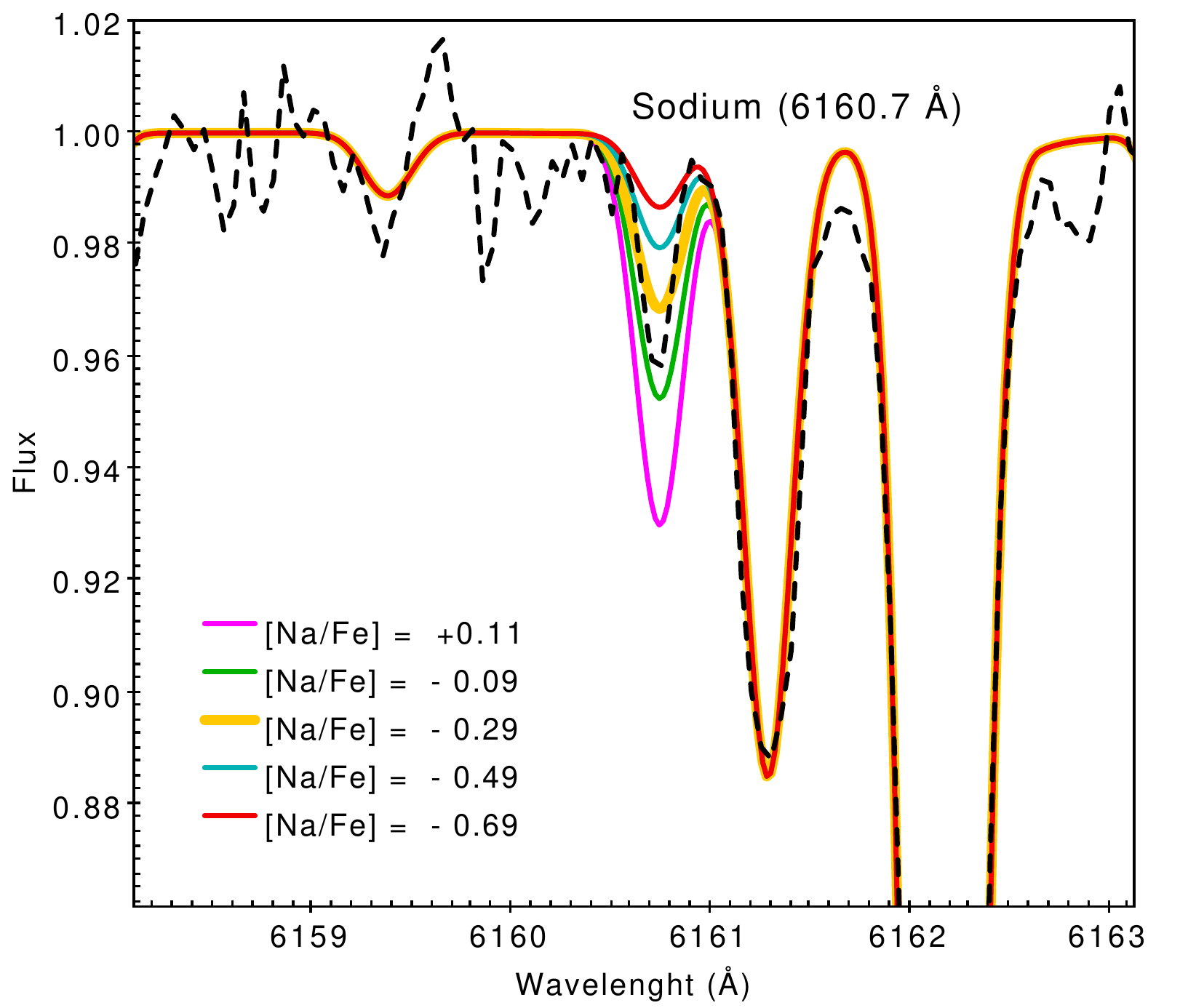}
\caption{\label{fig7} Spectrum-synthesis in the line at 6160.7 \AA \ of the star 12911. Five different synthetic spectra appear as coloured lines. The best-fitting among these is shown as a broader yellow line ([Na/Fe]=-0.29).}
\end{figure} 

We used the following lines for iron: (6136 \AA, 6191 \AA, 6213 \AA, 6252 \AA, 6322 \AA, 6335 \AA, 6336 \AA, and 6393 \AA), while for Na we used the line at 6160.7 \AA \ since that at 6154.2 \AA \ was too weak. The FWHM to be used for the spectrum-synthesis of the Na line was determined by the comparison of the synthetic spectra with nearby strong and well defined Ca and Fe lines. An example of the spectrum-synthesis applied to the Na line is shown in Fig. \ref{fig7}.
For some targets we could estimate only upper limits. The adopted solar abundances were log$\epsilon$(Fe)=7.50 and log$\epsilon$(Na)=6.32 (V13).
Our mean abundances for each star are represented in table \ref{abundances}. Na is known to be affected by departure from LTE. In this paper we did not apply any NLTE correction since our analysis is based on the relative Na abundance of stars that have roughly the same atmospheric parameters and so the same NLTE corrections for sodium. In any case, according to the INSPECT database \footnote{http://www.inspect-stars.com/}, the Na NLTE correction for our abundances is of the order of -0.20 dex for all our targets.
\begin{table}
\centering
 \begin{tabular}{c c c c} 
 \hline
 ID & T$_{\text{eff}}$ & [Fe/H] & [Na/Fe]\\ [0.5ex] 
 \hline
  11012 & 4978 & -1.42 & <-0.20\\
  11579 & 4856 & -1.44 & -0.39\\
  12911 & 4382 & -1.53 & -0.29\\
  14650 & 4843 & -1.43 & <-0.29\\
  14861 & 4888 & -1.41 & <-0.01\\
  15108 & 4891 & -1.47 & < 0.05\\
  15225 & 4920 & -1.55 & < 0.03\\
  15502 & 4761 & -1.47 & -0.31 \\
  15985 & 4723 & -1.46 & <-0.16\\
  16174 & 4928 & -1.42 & -0.37 \\
  16394 & 4501 & -1.53 & -0.28 \\
5015399 & 4865 & -1.44 & -0.44 \\
5016747 & 4856 & -1.46 & -0.41 \\ [1ex] 
 \hline
\end{tabular}
\caption{Table with abundance values. The order of the columns are: Star ID, effective temperature, metallicity and Na abundance. In some cases we could establish only upper limits for Na abundances.}
\label{abundances}
\end{table}
\begin{figure}
\includegraphics[width=0.5\textwidth]{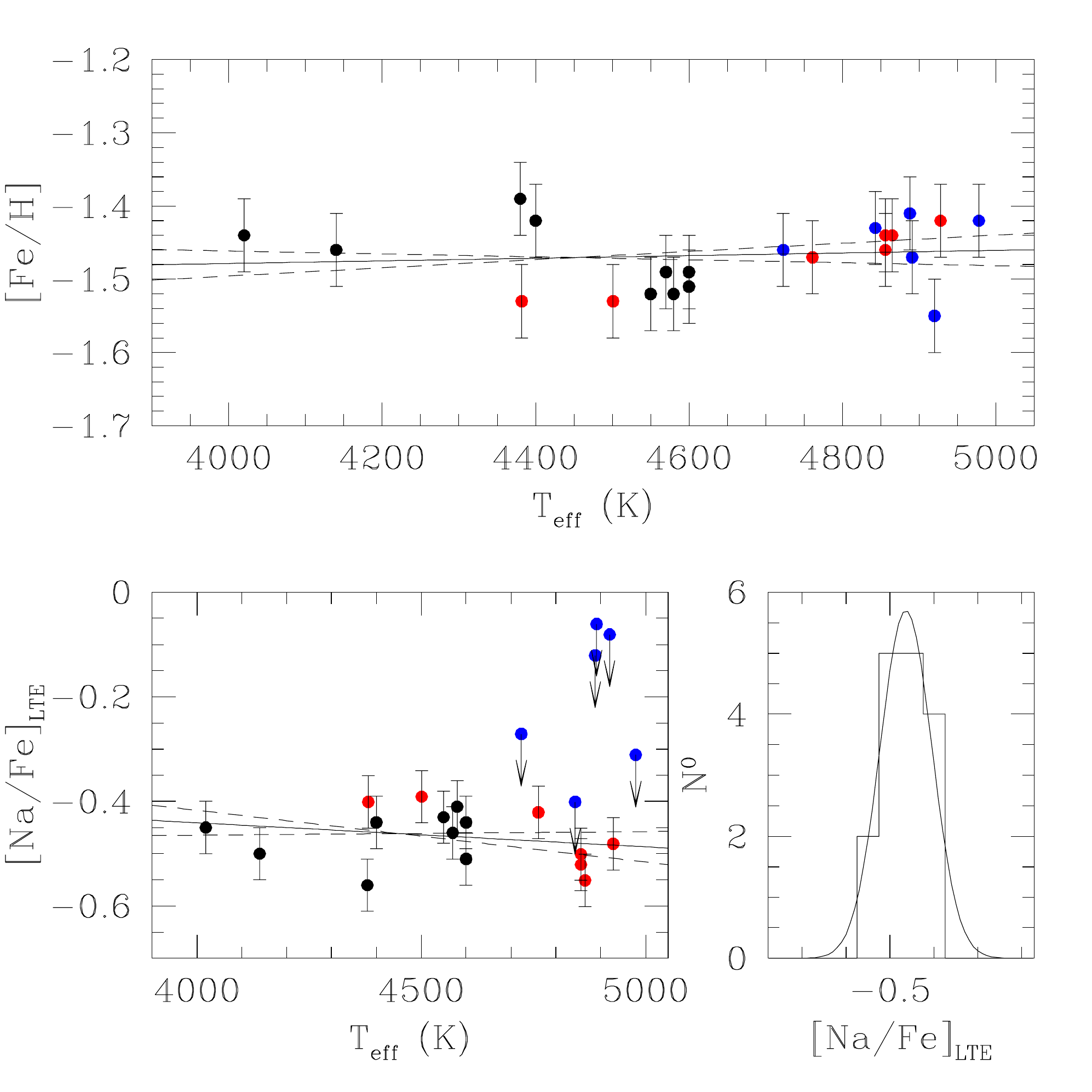}
\caption{\label{eps} Upper panel: [Fe/H] abundance as a function of the temperature. Lower panel(Left): [Na/Fe]$_{\text{LTE}}$ abundances as a function of the temperature. Lower panel(Right): [Na/Fe]$_{\text{LTE}}$ distribution.}
\end{figure} 

Fig. \ref{eps} report the present results together with those of V13. In the upper panel we report the [Fe/H] abundance as a function of the temperature. We can see that there is no trend in spite that the two sets of abundances were obtained using different spectrographs and methods. Also the linear fit is compatible with a flat trend with 1 $\sigma$. 
Combining the two databases we obtained a mean iron abundance of
$$[Fe/H]=-1.47\pm0.01$$

and

$$\sigma_{[Fe/H]}=0.05\pm0.01$$
In the lower panels we report the [Na/Fe]$_{\text{LTE}}$ abundances as a function of the temperature (left panel) and the [Na/Fe]$_{\text{LTE}}$ distribution (right pannel). In this case GIRAFFE data have a systematic shift of +0.11 dex (targets with upper limits were not considered for the comparison). The cause of this systematics is probably due to the fact that in V13 we used the four Na lines at 5682.6 \AA , 5688.2 \AA, 6154.2 \AA \ and 6160.7 \AA \ as Na abundance indicator while here we could use only that at 6160.7 \AA. Also systematics due to some effect related to the spectrograph not well removed during the reduction procedure cannot be ruled out such as scattered light. We applied a correction of -0.11 dex to the Na abundances obtained from GIRAFFE data.
We found a mean Na LTE abundance of:
$$[Na/Fe]=-0.47\pm0.01 $$
The Na distribution histogram is very narrow with a r.m.s. of:
$$\sigma_{[Na/Fe]}=0.06\pm0.01$$
The typical internal error on the fit for our T$_{\text{eff}}$ is 10-20 K, while the errors on log(g) and v$_\text{t}$ are below 0.1 dex and 0.05 km/s respectively. If we apply the same procedure described in V13 for the error calculation we obtain:
$$\sigma_{TOT} (Na)=0.05$$
We underline the fact that in our case the  observational error is dominated by the S/N of the spectra.
Comparing this value with the r.m.s. of the Na distribution histogram we can confirm the result by V13 that Rup106 does not host multiple stellar populations.


\section{Photometric Analysis}
\begin{figure}
\includegraphics[width=0.5\textwidth]{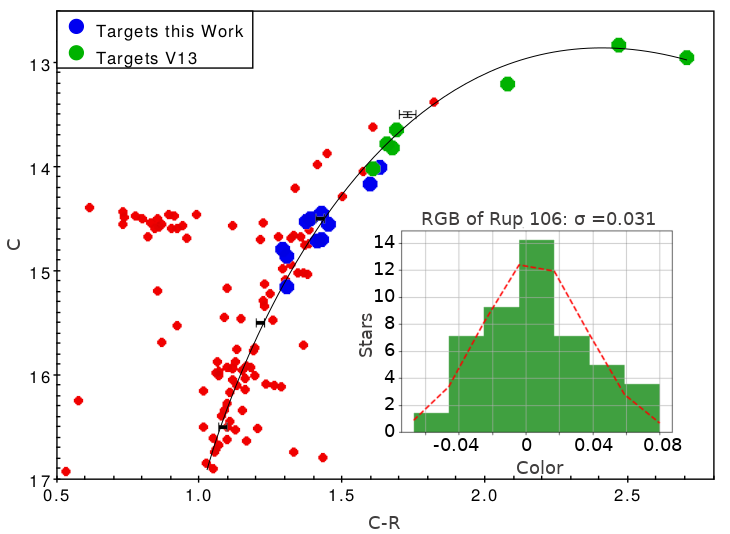}
\caption{\label{c-r} CMD using the Washington filter C to distinguish the presence of MPs, the spread of the Targets along the RGB indicate the presence of only one population. A fiducial has been adjusted, color differences of stars from the fiducial have been normalized in a histogram and the best-fitting gaussian has been derived.}
\end{figure}

As mentioned, the Washington filter C has proved to be useful to detect MPs due to the fact that it covers CN and NH bands \citep{Canterna1976}. Figure \ref{c-r} shows a CMD obtained using the Washington Filter C combined with the R$_{\text{KC}}$ filter, limited to the part of the RGB where our targets lie.
  If Rup106 had more than one stellar population we should observe one of the following effects: \\
 a) A split in the RGB \citep{Cummings2014}. In this case all the targets would lie in one of the RGBs. \\
 b) A spread in the RGB caused by the chemical differences between the populations and larger than the photometric errors.\\
 Figure \ref{c-r} shows instead that the RGB of Rup106 is very narrow and that the spread in color is compatible with the errors. 
 
 A fiducial curve (defined as the highest density locus of stars along the RGB) has been fitted along the RGB in C-R vs C (the black curve in \ref{c-r}). We then measured the color difference of the stars from the fiducial and build a distribution histogram of this value. We then derived the best-fitting gaussian for the histogram and got $\sigma$ = 0.031 $\pm$ 0.003, about 1.5 times the median error in the C-R color for the RGB that is $\sigma_{\text{C-R}}$ = 0.02 $\pm$ 0.01 
 obtained calculating the square root of the sum of the squares of the errors of each filter. The $\sigma$ value we found is very likely an upper limit and not the intrinsic width of the RGB since the field contamination cannot be fully removed because of the superposition of the cluster with the field in the proper motion space (see Fig. \ref{pm}). For this reason we conclude that the width of the Rup106 RGB is fully explained by the photometric errors and it does not require the presence of multiple stellar populations.


\section{The Orbit}

We used the \textsc{gravpot16}\footnote{https://gravpot.utinam.cnrs.fr/} model (Fern\'andez-Trincado et al. in preparation) to study the orbital elements (eccentricity, apo-/peri-galactocentric distance, the characteristic orbital energy, and the orbital Jacobi constant) of Rup106. Since V13 already show that this cluster has an extragalactic origin based on its Na and $\alpha$-element abundances, the aim is to find the Halo structure Rup106 is associated with. 

\begin{figure*}
\includegraphics[width=\textwidth]{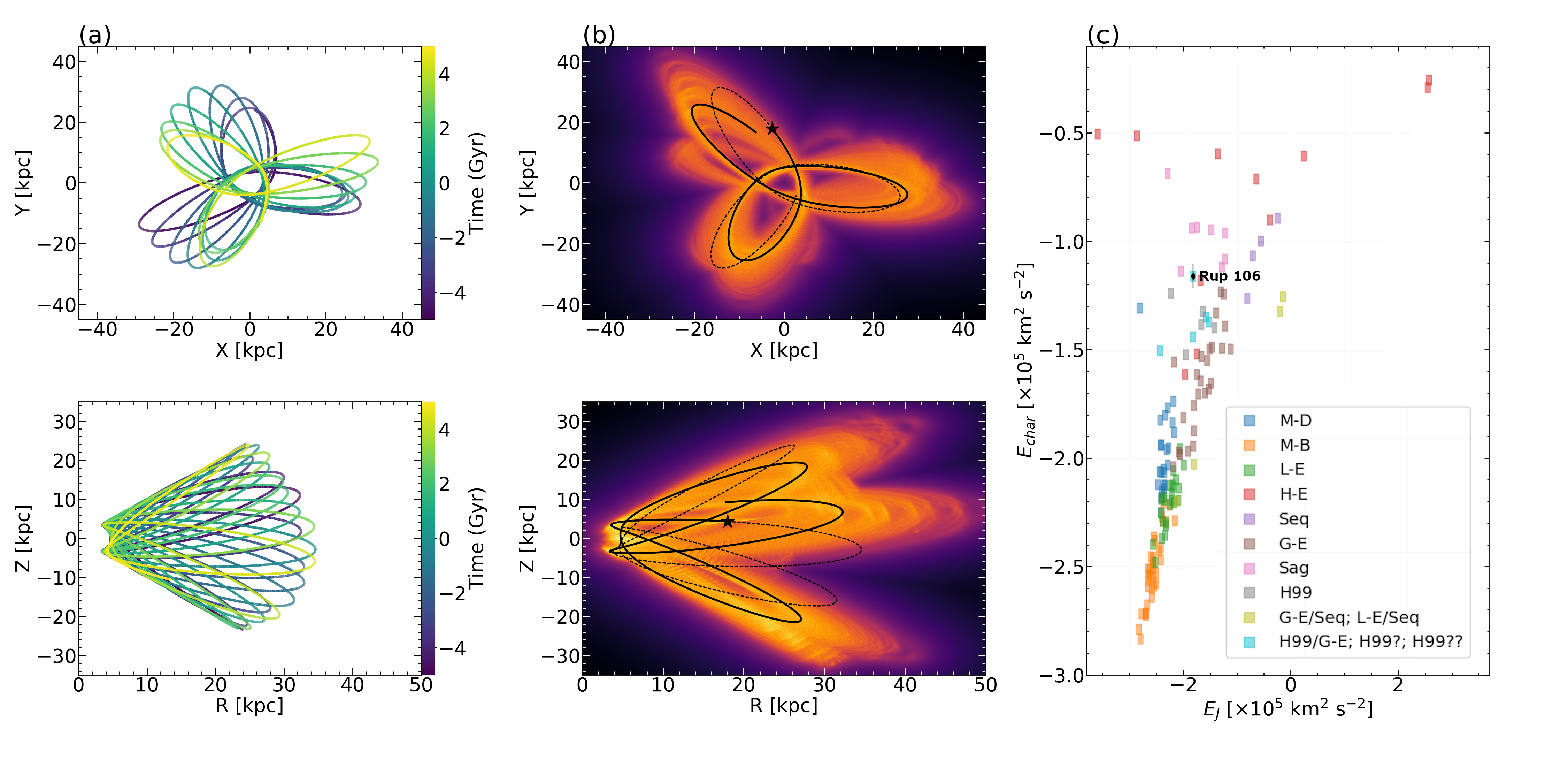}
\caption{Panels (a) and (b) shows the equatorial and meridional Galactic planes in the inertial frame, time-integrated forward/backward over 5 Gyr. Panel (b) show the probability density, with yellow and orange colors corresponding to more probable regions of the space, which are crossed more frequently by the simulated orbits. The black solid and dashed line show the forward and backward orbital path of Rup106 over 1 Gyr, for guidance. Panel (c) show the Characteristic orbital energy ((E$_{\text{max}}$ + E$_{\text{min}})$/2) versus the orbital Jacobi constant (E$_{\text{J}}$) in the non-inertial reference frame where the bar is at rest. Other Galactic GCs associated with different progenitors from \citep{Massari2019} are shown for comparison. The black dot with error bars refers to Rup106 analyzed in this work.}\label{orbits}
\end{figure*} 

The \textsc{gravpot16} is composed of a massive ($\sim$1.1$\times10^{10}$ M$_{\odot}$) 'boxy/peanut' bar/bulge structure accompanied by multiple stellar discs whose profiles mimic to that of the Besan\c{c}on Galaxy model \citep{Robin2003, Robin2014}. For the orbit computations, we adopt the same model configuration and Sun's positions and velocity as in \citet{Fernandez2020}, except for the bar patterns speed, which we adopt the recommended value of 41 km s$^{\text{-1}}$ kpc $^{\text{-1}}$ (see e.g., \citet{Sanders2019}). 
We integrated hundred thousand orbits by  adopting  a  simple  Monte Carlo  approach  which  considers the errors in the observables as 1-$\sigma$ variations over a 5 Gyr timespan toward the past (backward) and future (forward) by adopting the observables with their respective errors from \citet{Baumgardt2019}

\begin{center}
    
RA: 189.6675$\degree$\\
DEC: -51.150277$\degree$\\
d = 21.2 $\pm$ 2.12 kpc\\
RV$_{\text{Helio}}$: -38.36 $\pm$ 0.26 km s$^{\text{-1}}$\\
pm$_{\text{RA}}$: -1.25 $\pm$ 0.01 mas yr$^{\text{-1}}$\\
pm$_{\text{DEC}}$: 0.39 $\pm$ 0.01 mas yr$^{\text{-1}}$\\
\end{center}

Figure \ref{orbits} show the resulting orbits of Rup106 on the equatorial and meridional Galactic planes in the inertial frame. The top and bottom panel in Figure \ref{orbits}(a) show the predicted orbit of Rup106 without considering the errors in the observable, while the the top and bottom panel Figure \ref{orbits}(b) show the resulting ensemble of orbits from our Monte Carlo approach, which consider the errors in the observable. The yellow and orange colors correspond to more probable regions of the space, which are crossed more frequently by the simulated orbits, while the black solid and dashed line show the forward and backward orbital path of Rup106 over 1 Gyr for guidance.    

 Figures \ref{orbits}(a) and (b) reveals that Rup106 lies on a radial and highly eccentric ($>$0.81 $\pm$ 0.01) halo-like orbit with rather higher excursions above the Galactic plane ($\sim$23.6 $\pm$ 3.2 kpc). The perigalactocentric (r$_{\text{min}}$) and apogalactocentric (r$_{\text{max}}$) distance of Rup106 is $\sim$3.4 $\pm$ 0.5 kpc and $\sim$32.7 $\pm$ 3.7 kpc, respectively, placing the cluster well within the inner halo of the Milky Way, but located beyond of the bulge/bar region. In addition, using a slightly different angular velocity for the bar ($\pm$10 km s$^{\text{-1}}$ kpc$^{\text{-1}}$) does not change significantly our conclusions, and returns orbits in which the cluster is confined to the inner halo.

 It is important to note that unlike \citet{Baumgardt2019}, our orbit computations are based  in  a  realistic (as  far  as  possible) barred Milky Way  model, which may affect the orbital path of Rup106, as the cluster orbit has close approaches ($\sim$3 kpc) to the `bulge/bar' region, where the strength of the 'bar' structure is important. 

Figure \ref{orbits}c show the Characteristic orbital energy (E$_{\text{char}}$ = (E$_{\text{max}}$ + E$_{\text{min}}$)/2) versus the orbital Jacobi constant (E$_{\text{J}}$) in the non-inertial reference frame where the bar is at rest, as defined in \citet{Moreno2015} and \citet{Fernandez2020}. This plane reveals that the orbit of Rup106 lies in the boundary between three groups of GCs, e.g., those in the High-Energy group (H-E), the group dominated by Helmi-Stream (H99), and the group associated with the Sagittarius dwarf galaxy (Sgr) \citep[see e.g.,]{Massari2019}. For this reason, based only in the dynamical configuration of Rup106 there is not a clear association with any of the proposed progenitors in the Milky Way.

V13 concluded that the very low Na and $\alpha$-element abundances of Rup106 only match those of the Magellanic Clouds and of the Sagittarius Galaxy. Combining our results with these conclusions we could determine that the progenitor of Rup106 is the Sagittarius dwarf galaxy, adding evidence that does not contradict the results from 
\newline
\citet{Bellazzini2003} and are in line with those from \citet{Sbordone2005}, however it still contradicts the conclusion of \citet{Law2010} who did not find significant evidence for association with any wrap of the Sgr arms, leaving the discussion opened again.

In spite of the fact that the progenitor of Rup106 is not clear, \citet{Massari2019} and \citet{Bajkova2020} define it as a potential Helmi-Stream (H99) member. It is worth mentioning that all the other members of this group possess MPs with the exception of E3 (although only classified as H99 by \citet{Massari2019}). This cluster was studied in \citet{Salinas2015} and \citet{Monaco2018} analysing 23 RGB members with low resolution spectroscopy and 4 RGB with high resolution spectroscopy, respectively. Both studies conclude that there is no evidence of MPs in such cluster.

In addition, \citet{Bastian2018} named other 3 SSP GCs: Terzan 7, Pal 12 and Pal 3. The first two GCs are Sgr members and the last belongs to the H-E group. This would indicate that all the SSP and potential SSP GCs have an extra galactic origin.
 

\section{Conclusions}
In this article we have derived atmospheric parameters and chemical abundances for Fe and Na for 13 RGB stars of the GC Rup106 using FLAMES-GIRAFFE data. The abundance results have been compared with \citet{Villanova2013}.A photometric analysis with images taken from the 1-meter Swope Telescope was done as a complement to the spectroscopic results. For this purpose we studied the broadening of the RGB of Rup106 in the CMD using a filter sensible to the presence of MPs. Finally, we studied the orbit of the cluster and tried to associate it to some known Halo stream.
\newline
From these studies we can conclude the following:

1) Rup106 has [Fe/H] = -1.47 $\pm$ 0.01 and [Na/Fe]$_{\text{LTE}}$ = -0.47 $\pm$ 0.01. The [Fe/H] is in good agreement with V13, and the Na abundances confirm that the cluster does not have multiple stellar populations.

2) The RGB$_{\text{Broadening}}$/RGB$_{\text{error}}$ ratio in the Color-Magnitude Diagram of Rup106 is 1.5. This indicates that although there is a difference between both values, it is no enough large to contradict the spectroscopic result.

3) The orbital analysis indicates that Rup106 is confined to the halo, while the orbital energy puts Rup106 among the High-Energy group (H-E), Helmi-Stream (H99), and Sagittarius dwarf galaxy (Sgr).

It is interesting to note that combining our results concerning the orbits with the analysis from V13 we could propose the Sagittarius dwarf galaxy as the progenitor of Rup106. However, the work made in \citet{Law2010} indicates the opposite, leaving the question open.



\section*{Acknowledgements}
We thank the referee for helpful comments that greatly improved the paper.
SV gratefully acknowledges the support provided by Fondecyt regular n. 1170518.\\
J.G.F-T is supported by FONDECYT No. 3180210.\\
HF acknowledges financial support from Agencia Nacional de Investigacion y Desarrollo (ANID) grant 21181653.\\
CM thanks the support provided by  FONDECYT No. 1181797 and  from the Chilean Centro de Excelencia en Astrof\'isica y Tecnolog\'ias Afines (CATA) BASAL grant AFB-170002.
\section*{Data Availability}
This work has made use of data from the European Space Agency (ESA) mission
{\it Gaia} (\url{https://www.cosmos.esa.int/gaia}), processed by the {\it Gaia}
Data Processing and Analysis Consortium (DPAC, \url{https://www.cosmos.esa.int/web/gaia/dpac/consortium}). Funding for the DPAC has been provided by national institutions, in particular the institutions
participating in the {\it Gaia} Multilateral Agreement.

This work has made use of data from the ACS GC Survey available from the ACS GC Treasury database.

The spectroscopic raw data analysed here was observed under the Programme 098.D-0227(A) and can be obtained from the ESO Science archive.

The Photometric raw data analysed in this article will be shared on reasonable request to the corresponding author (HF).







\bsp	
\label{lastpage}
\end{document}